\documentclass[twocolumn]{aastex62}

\shorttitle{W0808}
\shortauthors{Flaherty et al.}

\begin{document}

\title{The Planet Formation Potential Around a 45 Myr old Accreting M Dwarf}

\correspondingauthor{Kevin Flaherty}
\email{kmf4@williams.edu}

\author[0000-0003-2657-1314]{Kevin Flaherty}
\affil{Department of Astronomy and Department of Physics\\
Williams College, Williamstown, MA 01267 USA}
\affil{Van Vleck Observatory, Astronomy Department, Wesleyan University \\
96 Foss Hill Drive \\
Middletown, CT 06459, USA}

\author{A. Meredith Hughes}
\affil{Van Vleck Observatory, Astronomy Department, Wesleyan University \\
96 Foss Hill Drive \\
Middletown, CT 06459, USA}

\author{Eric E. Mamajek}
\affil{Jet Propulsion Laboratory\\
California Institute of Technology\\
4800 Oak Grove Dr., Pasadena, CA 91109, USA}
\affil{Department of Physics and Astronomy, University of Rochester\\
500 Wilson Blvd., Rochester, NY 14627, USA}

\author{Simon J. Murphy}
\affil{School of Physical, Environment and Mathematical Sciences \\
University of New South Wales Canberra, ACT 2600, Australia}

\begin{abstract}
Debris disk detections around M dwarfs are rare, and so far no gas emission has been detected from an M dwarf debris disk. This makes the 45 Myr old M dwarf WISEJ080822.18-644357.3 a bit of a curiosity; it has a strong infrared excess at an age beyond the lifetime of a typical planet-forming disk, and also exhibits broad H$\alpha$ emission consistent with active accretion from a gaseous disk. To better understand the cold gas and dust properties of this system, we obtained ALMA observations of the 1.3mm continuum and the CO/$^{13}$CO/C$^{18}$O J=2-1 emission lines. No cold CO gas is detected from this system, ruling out a gas-rich protoplanetary disk.  Unresolved dust continuum emission is detected at a flux of 198$\pm$15 $\mu$Jy, consistent with 0.057$\pm$0.006 M$_{\oplus}$ worth of optically thin dust, and consistent with being generated through a collisional cascade induced by large bodies at radii $<$16 au.  With a sufficiently strong stellar wind, dust grains released in the outer disk can migrate inwards via PR drag, potentially serving as a source of grains for the strong infrared excess.

\end{abstract}



\section{Introduction}
Debris disks, created through the collisional grinding of planetesimals, are an important signpost of (often) unseen planetary systems. While debris disks have been found around $\sim$20\%\ of AFGK stars \citep{mat14,wya15,hug18}, they are detected much less frequently around M dwarfs. Infrared surveys put the occurrence of debris disks around field M dwarfs at $<$1.4\% \citep{pla05,pla09,ave12}. This dearth of collisional debris stands in contrast to the frequency of planets around M stars, with observational studies finding more than one planet per M dwarf \citep[e.g.][]{dre13,dre15,mul15,cla16}. 

M dwarf debris disk detection rates become more substantial when moving toward younger sources and/or longer wavelengths. \citet{for08} find that 4.3\% of M dwarfs in the 30-40 Myr old NGC 2547 cluster have a 24$\micron$ excess indicative of a debris disk, while \citet{bin17} measure a significant 22$\micron$ excess among 13$\pm$5\% of M dwarfs younger than 30 Myr. Sub-mm surveys have detected cold dust emission from three sources, out of 35 surveyed \citep{liu04,les06}, consistent with a 13\%\ detection rate. Two of the most well studied M dwarf debris disks, TWA 7 \citep{low05,mat07,olo18} and AU Mic \citep{kal04,mac13,dal18}, have ages of 10 Myr and 24 Myr respectively \citep{bell15}, consistent with the enhanced detectability of M dwarf debris disks at younger ages. 

Recently the 45 Myr old M dwarf WISEJ080822.18-644357.3 (W0808 hereafter) was discovered to have an infrared excess consistent with a debris disk \citep{sil16}, adding to the catalog of young M dwarfs with debris disks. Surprisingly, W0808 was subsequently found to have strong, and variable, H$\alpha$ emission consistent with the accretion of gas onto the stellar surface \citep{mur18}. Many of the prior detections of gas in debris disks have been found among high mass systems \citep{hug18}, with none showing evidence of substantial accretion onto the stellar surface. In fact, W0808 is the third M dwarf near the hydrogen burning limit to be discovered to be accreting at an age of $\sim$40 Myr \citep{rei09,bou16}. 

The presence of gas at such a late stage has important implications for the ability to form planets. The diminished frequency of gas-giant planets around low mass stars \citep{joh07,joh10,bon13,cla14,bow15,cla16} has often been attributed to the long timescales needed to build up massive cores around low mass stars \citep{lau04,pay07}.  While planet-forming disks around M dwarfs are less massive than the disks around their higher mass cousins \citep{and13}, they have a longer dispersal timescale \citep{her05,car06,ken09,luh12,rib15}, which may prolong the planet formation epoch. And if substantial gas reservoirs are common at 45 Myr, this may allow for more gas-giant planet formation at these relative late ages. The H$\alpha$ emission indicates that some gas is present within the W0808 system, although it is not clear if there is enough to populate the envelope of a gas-giant planet. 

Fully understanding the role of gas and dust in planet formation around W0808 requires a more detailed understanding of the total gas and dust available within the system. To that end we have observed W0808 with ALMA in search of a cold gas and dust reservoir. In section 2 we discuss the observations while in section 3 we discuss the detection of cold dust, and the non-detection of cold CO gas, from around W0808. In section 4 we discuss the possible origin of the mm continuum emission.

\section{Data\label{data}}
ALMA observations were taken on January 18, 2018 as part of project 2017.1.01521.S (PI: K. Flaherty). W0808 was observed with 45 antennas with baselines ranging from 15m to 1.4km, a total integration time of 41 min, and a mean precipitable water vapor column of 2 mm. The phase center of the observations was 08:08:22.1478 -64:43:56.742 (ICRS), chosen to match the predicted position of the system based on the pre-GAIA proper motion. One spectral window was devoted to the continuum at 232 GHz, with a 2 GHz bandwidth, while the remaining three windows were centered on CO(2-1), $^{13}$CO(2-1), and C$^{18}$O(2-1) respectively with 480 channels of width 244.141 kHz (0.38 km s$^{-1}$). The visibilities were calibrated using the standard ALMA calibration script, with J0635-7516 used for gain and bandpass calibration and J0904-5735 used for phase calibration. The phase center is shifted to match the stellar position at the epoch of the observations (08:08:22.178 -64:43:57.20) as predicted by GAIA DR2 data \citep{gaia18,lin18}.



\begin{figure*}[ht]
\includegraphics[scale=.5]{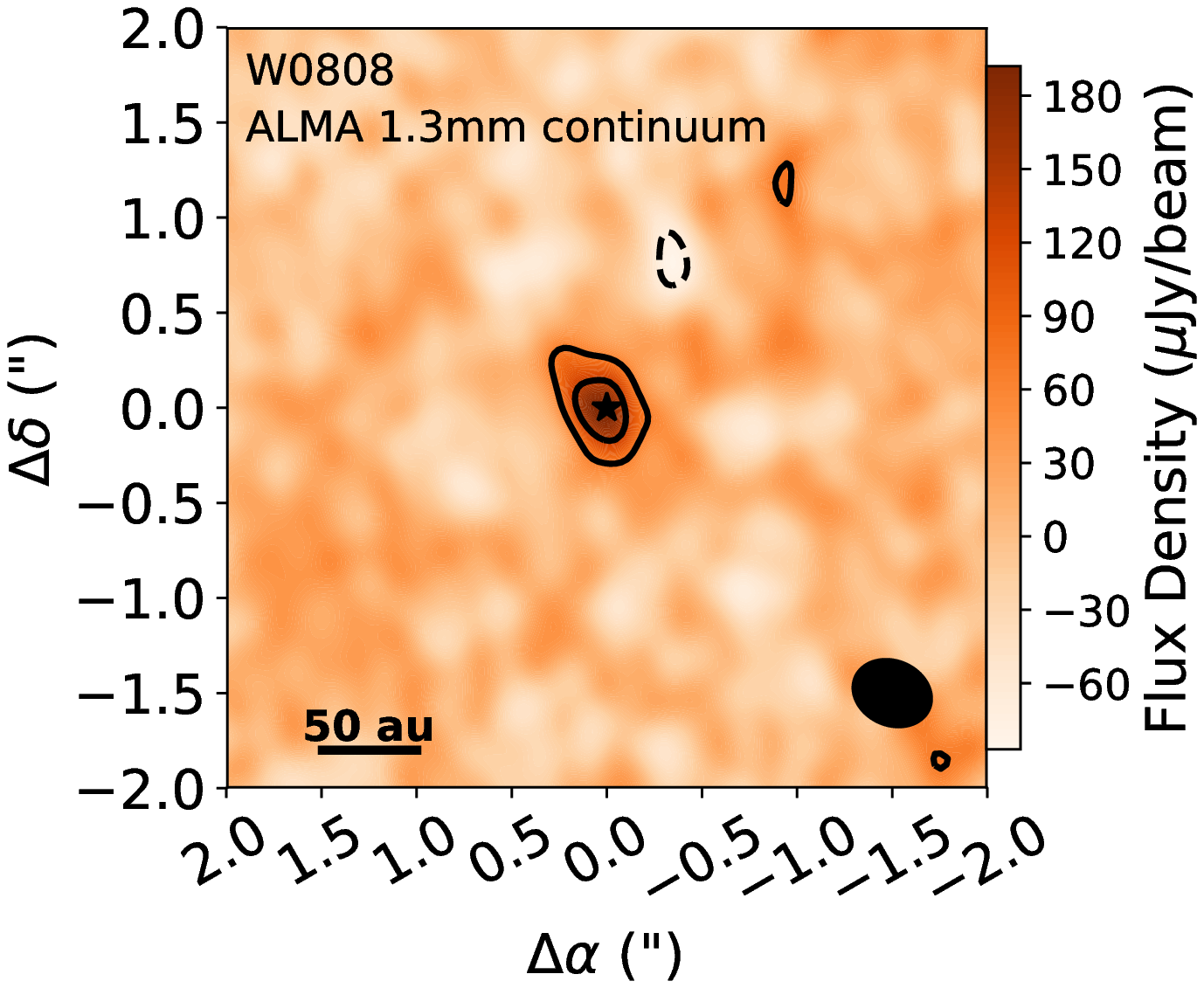}
\includegraphics[scale=.5]{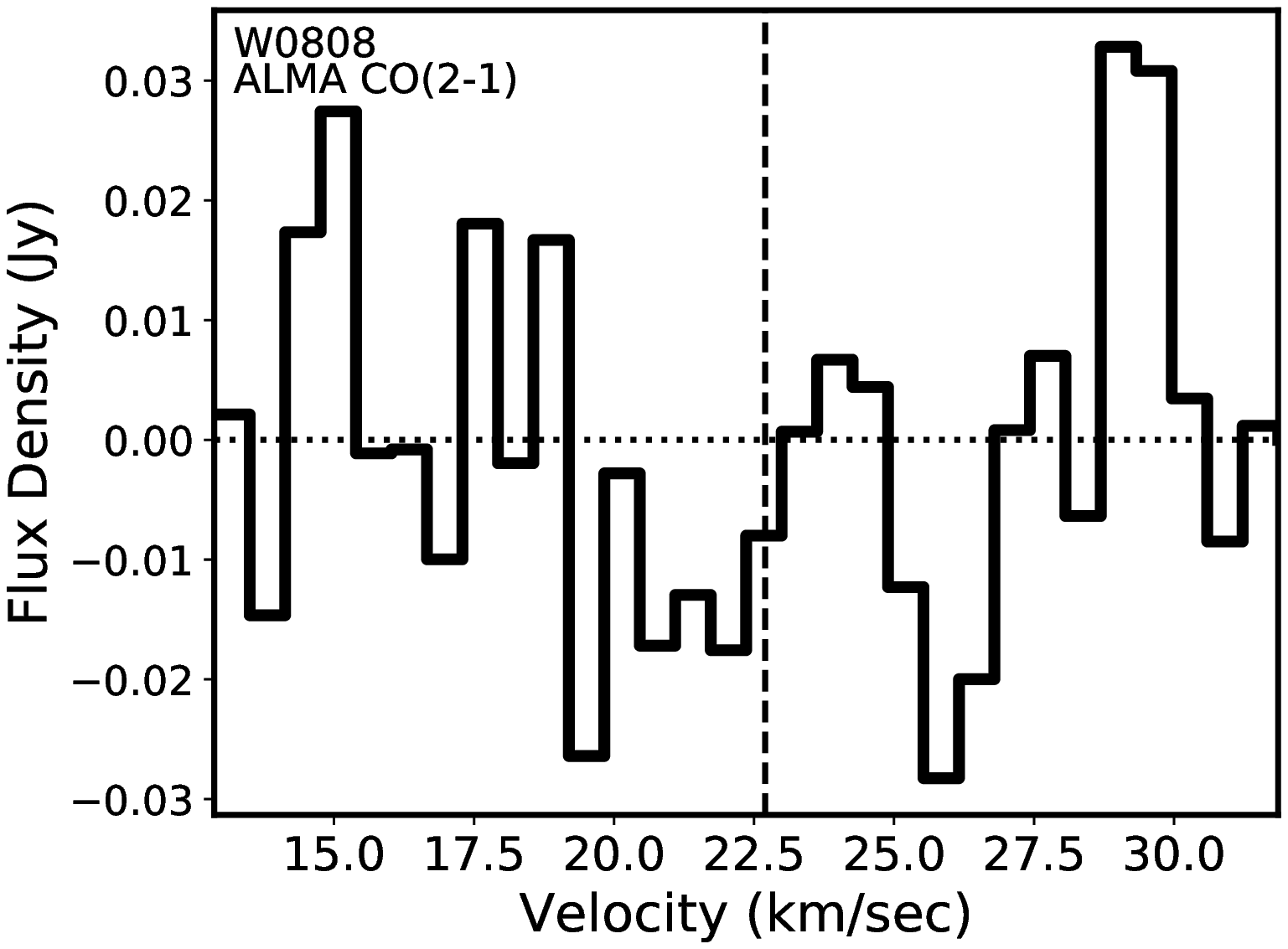}
\caption{{\it Left: } Dust emission is detected by ALMA at a wavelength of 1.3mm around the M dwarf W0808. The dust emission is detected at a peak S/N$\sim$8 (contours at 3$\sigma$ and 5$\sigma$, where $\sigma$=23$\mu$Jy/beam) and is unresolved (synthesized beam is indicated in the lower right, with a 50 au scale bar in the lower left). 
{\it Right:} Integrated emission within a 2$\arcsec$ x 2$\arcsec$ box centered at the stellar position, around the CO(2-1) emission line. No significant emission CO(2-1) emission is detected at the expected radial velocity of the system (marked by a vertical dashed line). We also do not detect any significant $^{13}$CO(2-1) or C$^{18}$O(2-1) emission. \label{continuum}}
\end{figure*}

No significant CO(2-1), $^{13}$CO(2-1), or C$^{18}$O(2-1) emission is detected (Figure~\ref{continuum}) in maps generated with natural weighting and with an rms of 1.8 mJy/beam/channel. Given the lack of gas emission, we use all four bands to image the continuum, with the CASA {\tt\string clean} task using natural weighting, which results in a beam of $0\farcs41$x$0\farcs32$ and an rms of 23 $\mu$Jy/beam. The continuum emission is significantly detected with a peak flux of 194 $\mu$Jy/beam, corresponding to a $\sim$8$\sigma$ detection (Figure~\ref{continuum}). The emission appears unresolved and a point source fit to the visibilities derives a total flux of 198$\pm$15 $\mu$Jy. Stellar flares can contaminate sub-mm searches of debris disks around M dwarfs \citep{mac18}, but we find no evidence for a strong stellar flare in the amplitude time series. 


Stellar parameters for W0808 (Table~\ref{stellar_properties}) are taken from \citet{mur18}, with updates to the stellar position, distance, luminosity, and proper motion based on the GAIA second data release \citep{gaia16,gaia18,lin18}. With the updated position, parallax, and proper motion from GAIA DR2, the BANYAN $\Sigma$ tool \citep{gag18} returns a 95\%\ probability of membership in the 45$^{+11}_{-7}$ Myr moving group Carina \citep{bell15} confirming the relative youth of W0808, although we caution that the Carina group itself is defined using only a small handful of stars \citep[see discussion in][]{mur18}. 


\begin{deluxetable}{lc}
\tablecaption{Stellar Properties\label{stellar_properties}}
\tablehead{\colhead{Parameter} & \colhead{Value}}
\startdata
RA & 08:08:22.182133 $\pm$ 0.000002\tablenotemark{a}\\
Dec & -64:43:57.26075 $\pm$ 0.00003\tablenotemark{a}\\
$\mu_{\alpha}\cos\delta$ & -11.54$\pm$0.12 mas yr$^{-1}$\tablenotemark{a}\\
$\mu_{\delta}$ & 25.61$\pm$0.10 mas yr$^{-1}$\tablenotemark{a}\\
RV & 22.7$\pm$0.5 km s$^{-1}$\\
distance & 101.4$\pm$0.6\tablenotemark{a} pc\\
T$_{\rm eff}$ & 3050$\pm$100 K\\
$\log$(L/L$_{\odot}$) & -2.05$\pm$0.08\tablenotemark{b}\\ 
M$_{*}$ (M$_{\odot}$) & 0.16$^{0.03}_{-0.04}$\\
Age & 45$^{+11}_{-7}$ Myr\\
Spectral Type & M5\\
\enddata
\tablenotetext{a}{Taken from GAIA DR2 \citep{gaia18,lin18}. RA and Dec are the stellar positions at epoch J2015.5, the reference epoch of the GAIA data.}
\tablenotetext{b}{Adjusted from \citet{mur18} based on new GAIA DR2 distance.}
\tablecomments{Stellar properties of W0808, drawn from \citet{mur18}.}
\end{deluxetable}

\section{Results}
\subsection{Cold Dust Emission}
\begin{figure}
\includegraphics[scale=.5]{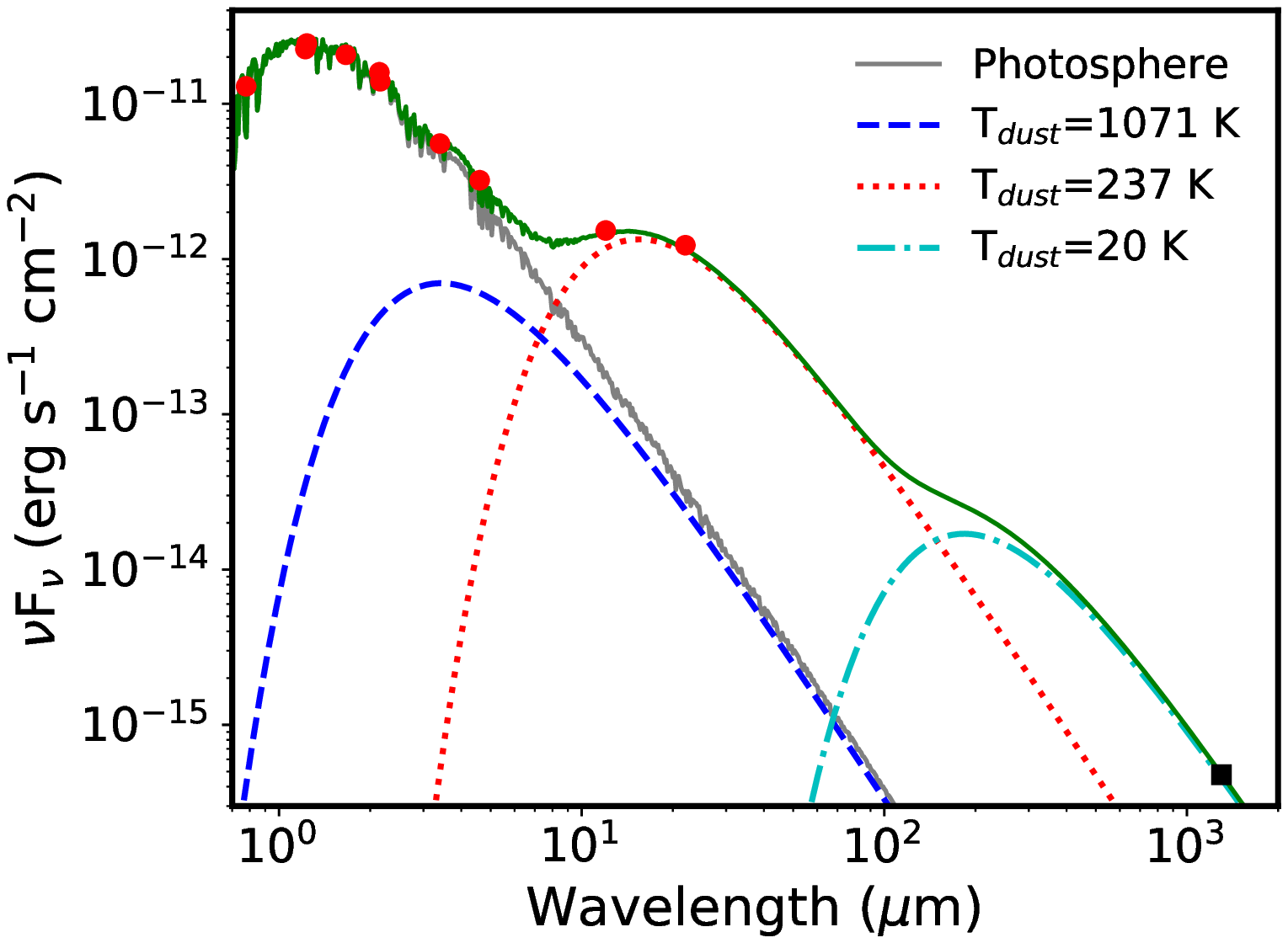}
\caption{Spectral energy distribution for W0808, with photometry drawn from DENIS \citep{epc99}, 2MASS \citep{cut03} and ALLWISE \citep{cut13}. A 3100 K BT-Settl \citep{all12,bar15} stellar photosphere is indicated by the grey line. The 1.3mm emission lies well above the predicted flux from the hot ($T_{\rm dust}$=1070 K, blue dashed line) and warm ($T_{\rm dust}$=237K, red dotted line) dust emission. An additional cold component, illustrated here with a 20 K blackbody (cyan dot-dashed line), is needed to reproduce the long-wavelength emission. \label{sed}}
\end{figure}

With ALMA we have detected significant dust emission from around the M dwarf W0808. At 198$\pm$14 $\mu$Jy the flux is much higher than the predicted 13$\mu$Jy flux from the warm dust emission, and is consistent with an additional component of cold dust (Figure~\ref{sed}). 

The emission is unresolved, which implies a disk radius $<$16 au at 101 pc. Blackbody grains 16 au from W0808 have a temperature of $T_{\rm dust}$= 20 K, consistent with the need of an additional cold component in the spectral energy distribution. Assuming that the dust emission is optically thin we can estimate the dust mass: 
\begin{equation}
M_d = \frac{F_{\nu}d^2}{\kappa_{\nu}B_{\nu}(T_{\rm dust})},
\end{equation}
where $F_{\nu}$ is the observed flux, $d$ is the distance to W0808, $\kappa_{\nu}$ is the dust opacity, and $T_d$ is the dust temperature. We derive a dust mass of 0.057$\pm$0.006 M$_{\oplus}$, assuming an opacity of 2.3 g cm$^{-2}$ at 1.3mm \citep{bec90}, and a dust temperature of 20 K. Given the unresolved nature of the emission, there is some uncertainty associated with this dust mass estimate. If the disk is smaller than 16 au in radius, the dust temperature is likely higher than assumed here, leading to a smaller dust mass. If the dust emission is compact enough to be optically thick then our estimate is a lower limit on the total dust mass. Debris disks are typically cold and optically thin, providing support for our initial assumptions, although additional observations at higher spatial resolution are needed to more accurately constrain the dust mass.

\begin{figure}
\center
\includegraphics[scale=.42]{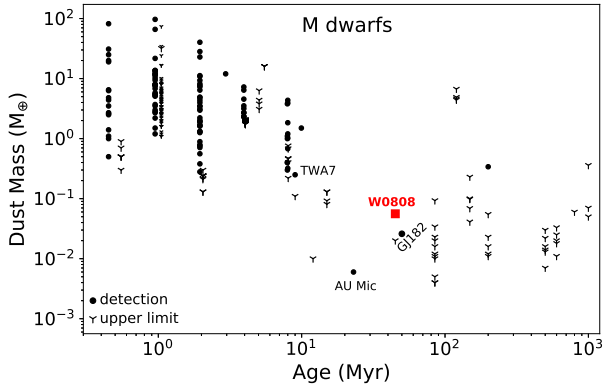}
\caption{Dust mass, as derived from mm continuum emission, as a function of age for disks around M dwarfs. Protoplanetary disks, with ages $\lesssim$ 10 Myr, have typical dust masses of 1-100 M$_{\oplus}$, while debris disk masses are typically less than 0.01-1M$_{\oplus}$. The small handful of detections (W0808 is marked with a red square) of dust masses among M dwarf debris disks suggest substantial evolution following the protoplanetary disk phase, although the many upper limits make it difficult to constrain the full extent of this evolution. \label{mass_age}}
\end{figure}

While the active accretion and strong infrared excess are suggestive of a gas-rich protoplanetary disk, the weak 1.3mm flux is more consistent with that of a debris disk. Figure~\ref{mass_age} shows dust mass, as derived from sub-mm emission, as a function of age for protoplanetary disks \citep{kle03,ric12,moh13,and13,ric13,ric14,car14,bro14,tes16,van16,ans16,ans17} and debris disks \citep{liu04,les06,mat07,les12,mat15,hol17} around M dwarfs. Protoplanetary disks have masses of 1-100 M$_{\oplus}$, with a factor of $\sim$5 decrease in the average mass between 1 and 10 Myr. Beyond 10 Myr, M dwarf debris disks are much less massive, with upper limits in the range of 0.01-1 M$_{\oplus}$. Many of these upper limits are for sources with ages $\gtrsim$100 Myr due to sensitivity constraints; sub-mm surveys of M dwarfs are limited to the nearest sources, which tend to be older. Among sources with ages similar to that of W0808, the dust mass around W0808 is larger than the 0.01 M$_{\oplus}$ dust mass around $\sim$20 Myr AU Mic \citep{mac13,mat15,dal18}, and the 0.03 M$_{\oplus}$ around the 50 Myr old GJ 182 \citep{liu04}, but smaller than the 0.25 M$_{\oplus}$ disk around the $\sim$10 Myr old TWA 7 \citep{hol17}.

\subsection{The lack of Gas Emission}
No significant CO(2-1), $^{13}$CO(2-1), or C$^{18}$O(2-1) emission is detected from the disk around W0808, suggesting that there is little cold gas in this system. Assuming the CO is confined to a uniform ring extending out to 16 au at a 45$^{\circ}$ inclination, which corresponds to a spatial coverage of $\sim$1 beam and a spectral coverage of $\sim$10 channels, we derive a 3$\sigma$ upper limit on the CO(2-1) flux of $<$10 mJy km s$^{-1}$. Only a handful of M dwarf protoplanetary disks have been detected in CO(2-1) or CO(3-2) emission, with typical flux levels, scaled to the distance of W0808, of 0.5-2.8 Jy km s$^{-1}$ \citep{ric12,ric14,van16}. The majority have CO flux upper limits corresponding to $\lesssim$80-1000 mJy km s$^{-1}$ at the distance of W0808 \citep{van16,ans17}. Molecular gas emission has only been detected within debris disks around more massive stars \citep{kos13,den14,mar16,lie16,whi16,hug17,moo17,mat17}, with typical CO(2-1) flux levels of $\sim$50 mJy km s$^{-1}$ to $\sim$20 Jy km s$^{-1}$, normalized to the distance of W0808, among objects with ages similar to W0808 \citep{hug18}.

Assuming local thermodynamic equilibrium (LTE), with an excitation temperature of 20 K, the flux limit corresponds to an upper limit on the cold CO gas mass of $<$5$\times10^{-6}$ M$_{\oplus}$. Outside of AU Mic \citep{dal18}, this represents the deepest limit on CO emission from around an M dwarf. We caution that full NLTE calculations are needed to accurately constrain the CO mass within these low density systems, which will likely result in a higher upper limit on the CO gas mass \citep{matra15}. 

While this limit falls well below typical protoplanetary disk gas masses \citep[e.g.][]{ans16}, it is more in line with predictions for gas released in the second-generation collisions that produce debris disks. \citet{kra17} present a model for CO emission from debris disks based on the collisional cascade of gas-rich planetesimals that predicts CO masses of 10$^{-6}$ - 10$^{-3}$ M$_{\oplus}$ for systems with measurable CO emission, all of which are more massive stellar systems than W0808. In the context of this debris disk model, the lack of detectable CO emission around W0808 is not surprising, although it does not account for the gas responsible for the H$\alpha$ emission. 

The active accretion, estimated to be $\sim$10$^{-10}$ M$_{\odot}$ yr$^{-1}$ based on the width of the H$\alpha$ emission \citep{mur18}, implies that some gas must be present in this system. One way for the gas to be hidden from our ALMA observations is if the gas is at a high temperature, reducing the population of CO molecules in the J=2 state. Based on the shape of the infrared excess, \citet{mur18} found evidence for two warm dust rings and for excitation temperatures corresponding to the dust temperatures of these two rings ($T_{\rm ex}$=1071 K, 237 K) the LTE upper limits on the CO mass are 1$\times$10$^{-4}$ M$_{\oplus}$ and 3$\times10^{-5}$ M$_{\oplus}$, respectively.  

Alternatively, the gas may have a low CO abundance, making our CO observations an inefficient method for tracing to gas within this disk. The lack of CO may be because the planetesimals responsible for driving the collisional cascade are e.g. H$_2$O-rich instead of CO-rich or because any CO released in the collisional cascade was rapidly destroyed by the radiation environment. Regarding the latter possibility, around massive A stars the UV radiation is strong enough to photodissociate CO within 10-100 years in low density debris disks \citep{mat18a}. The photodissociation of CO would release oxygen atoms into the disk, which may be the source of the detected [OI] 6300\AA\ emission \citep{mur18}. Recent surveys, with an eye towards the habitability of planets around M dwarfs, have greatly expanded the observations of M dwarf UV radiation, and have found UV fluxes orders of magnitude larger than that expected from the stellar photosphere \citep[e.g.][]{fra16,sch18}. This harsh radiation environment may lead to CO-poor gas around M stars.

To estimate the rate at which CO is photodissociated around W0808, we utilize the photodissociation rates of \citet{vis09}. For the stellar UV radiation field, we assume the ratio between the FUV and J-band flux densities is 4$\times$10$^{-5}$, the average observed value among 0.08-0.35 M$_{\odot}$, 45 Myr M dwarfs \citep{sch18}. We also include the interstellar radiation field as a Draine field flux that is independent of distance from the star. We find that close to the star ($<$ 10 au) the stellar radiation plays a large role in photodissociation, destroying CO molecules within $\sim$12 years. At larger radii the interstellar radiation field takes over, and the photodissociation timescale increases to $\sim$60 years at 20 au. This calculation assumes no self-shielding, and is likely a lower limit on the photodissociation timescale. We can estimate the role of self-shielding by assuming that a CO gas mass given by the LTE upper limit (5$\times$10$^{-6}$ M$_{\oplus}$) is spread evenly over a region with a diameter of 30 au. Under these conditions, the photo-dissociation timescale increases to 150-350 years between 10 and 20 au. Even with self-shielding the photo-dissociation timescale is much smaller than the age of the system, indicating that any primordial gas will be heavily depleted of CO. Any CO gas released in the collisions of gas-rich bodies will also be quickly destroyed without continuous replenishment.

\section{The Origin of the Cold Dust Emission}
\subsection{Planetesimal Collisions}
The lack of CO emission and the modest sub-mm emission are consistent with a debris disk in which dust grains are released in a collisional cascade induced by the collisions of km-sized planetesimals. The maximum radial extent of the disk around W0808, 16 au, is smaller than the $\sim$ 30 au debris disk radius predicted by an extrapolation of the stellar luminosity-belt radius correlation derived by \citet{mat18}. The belt radius is also smaller than the sizes of other resolved M dwarf debris disks;  Herschel emission is detected from 25 to 60 au around GJ 581 \citep{les12} while AU Mic has sub-mm emission detected out to $\sim$40 au \citep{mac13,dal18}. Similarly scattered light has been detected from around the M dwarf TWA 7 out to 25 au \citep{olo18}, around TWA 25 out to $\sim$80 au \citep{cho16}, and around GSC 07396-00759 out to $\sim$100 au \citep{sis18}. Given that observations of M dwarf debris disks are often limited by sensitivity, prior studies may have only sampled the brightest, and possibly most extended, systems. Protoplanetary disks around M dwarfs, from which the km-sized planetesimals are created, are typically unresolved in the sub-mm, limiting their radii to $\lesssim$30 au \citep{sch09,tes16,ric12,ric13,van16}, with only a handful of disks showing outer edges at 50-150 au \citep[e.g.][]{ric14}.

Once released into the disk, dust grains will evolve based on their interactions with stellar radiation and stellar winds. Unlike their higher mass cousins, radiation pressure does not play a strong role in dust dynamics around M dwarfs; instead it is interactions with stellar winds that will dominate \citep{pla05,str06,aug06,pla09,sch15}. Observations of the wind mass loss rate are fairly limited among M dwarfs, with typical values ranging from 0.005 $\dot{M_{\odot}}$ to $<$50 $\dot{M_{\odot}}$ \citep{woo04,vid17}, consistent with models of winds driven by cool stars \citep{cra11,joh15}, where $\dot{M_{\odot}}$ is the Solar wind mass loss rate ($\sim2\times10^{-14}$ M$_{\odot}$ yr$^{-1}$). Based on the prescriptions of \citet{str06} we can estimate the role of radiation/wind pressure and Poynting-Robertson drag on the dust grain dynamics. In the case of W0808, a very large wind ($\dot{M}>1000\dot{M_{\odot}}$) is needed to blow out micron-sized dust grains but stellar wind drag is more effective, with a wind $\gtrsim$$\dot{M_{\odot}}$ able to drag micron-sized grains inward toward the star within the age of the system. This indicates that any dust released in collisions in the outer disk will rapidly fill in the inner disk with small dust grains, rather than creating the large halos that are seen around more massive stars \citep{su09,su15}. 

Substantial inward migration via PR drag suggests a possible common origin for the sub-mm and infrared emission. Dust grains are released in collisions in the outer disk, where they are seen in the sub-mm, and migrate inwards towards the star until they heat up enough to produce the observed infrared emission. The warm dust temperature of $T_{\rm dust}$=237 K \citep{mur18} corresponds to a distance of 0.1 au from the central star; if the dust mass estimated from the sub-mm emission were to migrate inwards to this inner radius it would be highly optically thick ($\tau\sim30000$ at 12$\micron$ assuming $\kappa$(12$\micron$)=9$\times10^2$ cm$^2$ g$^{-1}$ \citep{dra03}), making it capable of producing the strong infrared emission. This scenario does rely on a modest stellar wind ($\gtrsim\dot{M_{\odot}}$) to migrate the dust inwards within the lifetime of W0808, and more detailed knowledge of the outer disk dust mass is needed to determine if inward migration can account for the infrared excess.

\subsection{Planetary-body collisions}
Another possible scenario is the recent collision of planetary bodies, generating a large amount of small dust grains at a highly localized region within the disk \citep{jac14}. Debris disks with large dust excesses (L$_{dust}$/L$_*$$>$10$^{-3}$) have often been found to also exhibit infrared variability \citep{men12,men14,men15}, likely a result of the recent collision that produces the observed debris. W0808 has both a large dust excess (L$_{IR}$/L$_*$$\sim$0.1) and infrared variability \citep{mur18}, consistent with such a scenario. The presence of silica emission \citep[e.g.][]{lis09}, created by the rapid cooling of molten rock released in the high energy impact \citep{joh12,joh14}, would provide confirmation of this scenario.

\subsection{Contamination from a background galaxy\label{galaxy}}
\begin{figure}
\includegraphics[scale=.4]{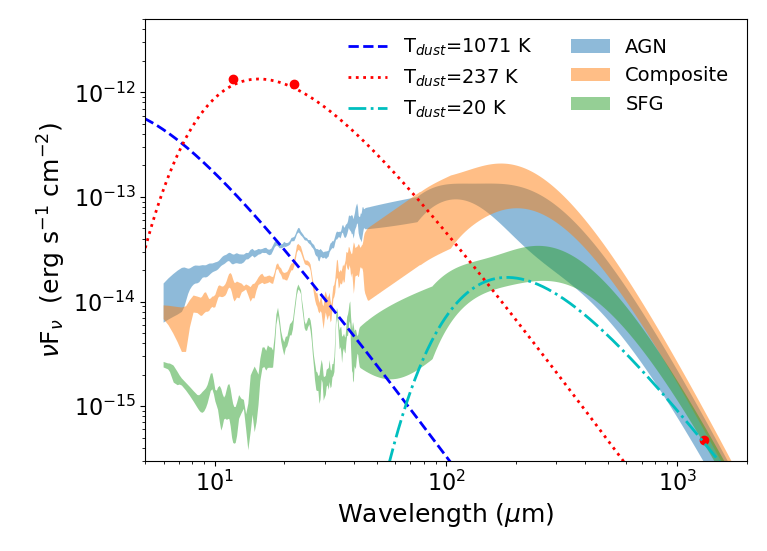}
\caption{SED of the long-wavelength emission, after subtracting off the contribution from the stellar photosphere. For comparison, we include SED templates for a z$\sim$2 active galactic nuclei (AGN), star-forming galaxy (SFG), and a composite SFG+AGN, taken from  \citet{kir15} (the bands demonstrate the range of observed fluxes from the galaxies used to generate the templates). The template SEDs were redshifted to z=2, and are normalized to the ALMA 1.3mm flux. The WISE infrared emission is orders of magnitude brighter than expected from a high-redshift galaxy, indicating that the excess emission is likely not from a high-redshift galaxy. \label{excess_sed}}
\end{figure}

One possibility for the origin of the ALMA emission is contamination from a background galaxy, a scenario that has been found in ALMA observations of the disks around HD 95086 \citep{su17} and TWA 7 \citep{bay18}. The shape of the SED can help distinguish between a disk and a background galaxy. Figure~\ref{excess_sed} shows the SED after subtracting off the stellar component, along with template SEDs for extragalactic sources \citep{kir15}, red-shifted to z=2 and scaled to the ALMA 1.3mm flux density. The infrared emission from around W0808 is 2-3 orders of magnitude brighter than that of a z=2 galaxy (similar results are found for galaxies at lower redshift), suggesting that the emission is not from a background galaxy. The $\sim$0$\farcs$21 offset between the ALMA and ALLWISE  \citep{cut13} emission positions is also inconsistent with the infrared and sub-mm emission both arising from a stationary background object, but is consistent with the GAIA measured proper motion. Broad H$\alpha$ emission and bright mid-IR fluxes are common among low-redshift AGN \citep[e.g.][]{sul00,sha18} but line widths among AGN are often much larger than the 300-400 km s$^{-1}$ observed around W0808, and the mid-IR flux from W0808 does not exhibit the rising power-law shape that is typical of AGN. 

Another scenario is that the H$\alpha$ and IR emission are from a disk surrounding W0808, while the ALMA emission is from an unrelated background galaxy. While we cannot rule out this possibility, it is highly unlikely; based on the Schechter function of \citet{car15}, there is a 0.01\%\ chance of detecting a background galaxy at 8$\sigma$ within 0$\farcs$5 of the stellar position. This suggests that the detected ALMA emission is most likely due to cold dust surrounding W0808.



\section{Conclusions}
We report the detection of unresolved 1.3 mm dust emission around the 45 Myr old accreting M dwarf W0808. While the active accretion and strong infrared excess are more similar to young planet-forming disks, the weak sub-mm emission, consistent with 0.057$\pm$0.006 M$_{\oplus}$ worth of optically thin dust, and lack of CO emission is more similar to debris disks. The sub-mm emission is unlikely to be due to the chance contamination from a background galaxy given the strong IR emission, and the small probability of a galaxy at the observed flux level in such proximity to the star. 

Around such a low-luminosity central source, any small grains released into the disk via collisions will not be blown away from the central star by radiation pressure, but are likely dragged inwards by PR drag. This raises the possibility that the dust grains are initially released in the outer disk, producing the observed sub-mm emission, and then migrate inwards, subsequently generating the observed infrared emission. Further observations, in particular constraints on the stellar wind mass loss rate, which sets the inward migration rate, are needed to fully constrain this model. The strong and variable infrared emission, in combination with weak sub-mm emission, may also be a sign of a recent collision of planet-sized bodies. Regardless of the exact physical origin of the dust, W0808 is likely in an early stage of planet formation and evolution as it transitions from a gas-rich planet-forming disk to a second-generation debris disk.



\acknowledgements
We thank the referee for their close reading of our manuscript. We thank Grant Kennedy, Jonathan Williams, and Nienke van der Marel for useful insight into possible explanations of the emission. E.E.M. acknowledges support from the NASA NExSS program. The work of K.M.F. was performed in part at the Aspen Center for Physics, which is supported by National Science Foundation grant PHY-1607611. K.M.F. and A.M.H. gratefully acknowledge support from NSF grant AST-1412647 This paper makes use of the following ALMA data: ADS/JAO.ALMA\#2017.1.01521.S. ALMA is a partnership of ESO (representing its member states), NSF (USA) and NINS (Japan), together with NRC (Canada), MOST and ASIAA (Taiwan), and KASI (Republic of Korea), in cooperation with the Republic of Chile, The Joint ALMA Observatory is operated by ESO, AUI/NRAO and NAOJ. The National Radio Astronomy Observatory is a facility of the National Science Foundation operated under cooperative agreement by Associated Universities, Inc. This work has made use of data from the European Space Agency (ESA) mission
{\it Gaia} (\url{https://www.cosmos.esa.int/gaia}), processed by the {\it Gaia}
Data Processing and Analysis Consortium (DPAC,
\url{https://www.cosmos.esa.int/web/gaia/dpac/consortium}). Funding for the DPAC
has been provided by national institutions, in particular the institutions
participating in the {\it Gaia} Multilateral Agreement. Part of this research was carried out at the Jet Propulsion Laboratory, California Institute of Technology, under a contract with the National Aeronautics and Space Administration.

\vspace{5mm}
\facilities{ALMA,GAIA}


\software{astropy \citep{astropy}, CASA \citep{mcm07}}


\begin{thebibliography}{}
\bibitem[Allard, et al.(2012)]{all12} Allard, F., Homeier, D. \& Freytag, B.\ 2012, Philosophical Transactions of the Royal Society of London Series A, 370, 2765.
\bibitem[Andrews, et al.(2013)]{and13} Andrews, S.~M., Rosenfeld, K.~A., Kraus, A.~L., et al.\ 2013, \apj, 771, 129.
\bibitem[Ansdell, et al.(2016)]{ans16} Ansdell, M., Williams, J.~P., van der Marel, N., et al.\ 2016, \apj, 828, 46.
\bibitem[Ansdell, et al.(2017)]{ans17} Ansdell, M., Williams, J.~P., Manara, C.~F., et al.\ 2017, \aj, 153, 240.
\bibitem[Astropy Collaboration et al.(2013)]{astropy} Astropy Collaboration, Robitaille, T.~P., Tollerud, E.~J., et al.\ 2013, \aap, 558, A33 
\bibitem[Augereau \& Beust(2006)]{aug06} Augereau, J.-C. \& Beust, H.\ 2006, \aap, 455, 987.
\bibitem[Avenhaus, et al.(2012)]{ave12} Avenhaus, H., Schmid, H.~M. \& Meyer, M.~R.\ 2012, \aap, 548, A105.
\bibitem[Ayliffe \& Bate(2009)]{ayl09} Ayliffe, B.~A. \& Bate, M.~R.\ 2009, \mnras, 397, 657.
\bibitem[Baraffe, et al.(2015)]{bar15} Baraffe, I., Homeier, D., Allard, F., et al.\ 2015, \aap, 577, A42.
\bibitem[Bayo, et al.(2018)]{bay18} Bayo, A., Olofsson, J., Gallardo, J., et al.\ 2018, ArXiv e-prints , arXiv:1806.09252.
\bibitem[Beckwith et al.(1990)]{bec90} Beckwith, S.~V.~W., Sargent, A.~I., Chini, R.~S., et al.\ 1990, \aj, 99, 924.
\bibitem[Bell et al.(2015)]{bell15} Bell, C.~P.~M., Mamajek, E.~E. \& Naylor, T.\ 2015, \mnras, 454, 593.
\bibitem[Binks \& Jeffries(2017)]{bin17} Binks, A.~S. \& Jeffries, R.~D.\ 2017, \mnras, 469, 579.
\bibitem[Bonfils, et al.(2013)]{bon13} Bonfils, X., Delfosse, X., Udry, S., et al.\ 2013, \aap, 549, A109.
\bibitem[Boucher et al.(2016)]{bou16} Boucher, A., Lafreni{\`e}re, D., Gagn{\'e}, J., et al.\ 2016, \apj, 832, 50
\bibitem[Bowler, et al.(2015a)]{bow15} Bowler, B.~P., Liu, M.~C., Shkolnik, E.~L., et al.\ 2015, The Astrophysical Journal Supplement Series, 216, 7. 
\bibitem[Broekhoven-Fiene, et al.(2014)]{bro14} Broekhoven-Fiene, H., Matthews, B., Duch{\^e}ne, G., et al.\ 2014, \apj, 789, 155.
\bibitem[Carniani, et al.(2015)]{car15} Carniani, S., Maiolino, R., De Zotti, G., et al.\ 2015, \aap, 584, A78.
\bibitem[Carpenter, et al.(2006)]{car06} Carpenter, J.~M., Mamajek, E.~E., Hillenbrand, L.~A., et al.\ 2006, \apj, 651, L49.
\bibitem[Carpenter, et al.(2014)]{car14} Carpenter, J.~M., Ricci, L. \& Isella, A.\ 2014, \apj, 787, 42.
\bibitem[Choquet, et al.(2016)]{cho16} Choquet, {\'E}., Perrin, M.~D., Chen, C.~H., et al.\ 2016, \apj, 817, L2.
\bibitem[Clanton \& Gaudi(2014)]{cla14} Clanton, C. \& Gaudi, B.~S.\ 2014, \apj, 791, 91.
\bibitem[Clanton \& Gaudi(2016)]{cla16} Clanton, C. \& Gaudi, B.~S.\ 2016, \apj, 819, 125.
\bibitem[Cranmer \& Saar(2011)]{cra11} Cranmer, S.~R. \& Saar, S.~H.\ 2011, \apj, 741, 54.
\bibitem[Cutri, et al.(2003)]{cut03} Cutri, R.~M., Skrutskie, M.~F., van Dyk, S., et al.\ 2003, VizieR Online Data Catalog , II/246.
\bibitem[Cutri \& et al.(2013)]{cut13} Cutri, R.~M. \& et al.\ 2013, VizieR Online Data Catalog , II/328.
\bibitem[Daley et al. (2018)]{dal18} Daley, C., Hughes, A.M., Carter, E., et al. in prep
\bibitem[Dent, et al.(2014)]{den14} Dent, W.~R.~F., Wyatt, M.~C., Roberge, A., et al.\ 2014, Science, 343, 1490.
\bibitem[Draine(2003)]{dra03} Draine, B.~T.\ 2003, Annual Review of Astronomy and Astrophysics, 41, 241.
\bibitem[Dressing \& Charbonneau(2013)]{dre13} Dressing, C.~D. \& Charbonneau, D.\ 2013, \apj, 767, 95.
\bibitem[Dressing \& Charbonneau(2015)]{dre15} Dressing, C.~D. \& Charbonneau, D.\ 2015, \apj, 807, 45.
\bibitem[Epchtein, et al.(1999)]{epc99} Epchtein, N., Deul, E., Derriere, S., et al.\ 1999, \aap, 349, 236.
\bibitem[Forbrich, et al.(2008)]{for08} Forbrich, J., Lada, C.~J., Muench, A.~A., et al.\ 2008, \apj, 687, 1107.
\bibitem[France, et al.(2016)]{fra16} France, K., Loyd, R.~O.~P., Youngblood, A., et al.\ 2016, \apj, 820, 89.
\bibitem[Gagn{\'e}, et al.(2018)]{gag18} Gagn{\'e}, J., Mamajek, E.~E., Malo, L., et al.\ 2018, \apj, 856, 23.
\bibitem[Gaia Collaboration, et al.(2016)]{gaia16} Gaia Collaboration, Prusti, T., de Bruijne, J.~H.~J., et al.\ 2016, \aap, 595, A1.
\bibitem[Gaia Collaboration et al.(2018)]{gaia18} Gaia Collaboration, Brown, A.~G.~A., Vallenari, A., et al.\ 2018, \aap, 616, A1.
\bibitem[Hern{\'a}ndez, et al.(2005)]{her05} Hern{\'a}ndez, J., Calvet, N., Hartmann, L., et al.\ 2005, \aj, 129, 856.
\bibitem[Hughes et al.(2017)]{hug17} Hughes, A.~M., Lieman-Sifry, J., Flaherty, K.~M., et al.\ 2017, \apj, 839, 86.
\bibitem[Hughes et al.(2018)]{hug18} Hughes, A.~M., Duch{\^e}ne, G., \& Matthews, B.~C.\ 2018, Annual Review of Astronomy and Astrophysics, 56, 541.
\bibitem[Holland, et al.(2017)]{hol17} Holland, W.~S., Matthews, B.~C., Kennedy, G.~M., et al.\ 2017, \mnras, 470, 3606.
\bibitem[Jackson, et al.(2014)]{jac14} Jackson, A.~P., Wyatt, M.~C., Bonsor, A., et al.\ 2014, \mnras, 440, 3757.
\bibitem[Johnson, \& Melosh(2012)]{joh12} Johnson, B.~C., \& Melosh, H.~J.\ 2012, \icarus, 217, 416.
\bibitem[Johnson, \& Melosh(2014)]{joh14} Johnson, B.~C., \& Melosh, H.~J.\ 2014, \icarus, 228, 347.
\bibitem[Johnstone, et al.(2015)]{joh15} Johnstone, C.~P., G{\"u}del, M., Brott, I., et al.\ 2015, \aap, 577, A28.
\bibitem[Johnson, et al.(2007)]{joh07} Johnson, J.~A., Butler, R.~P., Marcy, G.~W., et al.\ 2007, \apj, 670, 833.
\bibitem[Johnson, et al.(2010)]{joh10} Johnson, J.~A., Aller, K.~M., Howard, A.~W., et al.\ 2010, Publications of the Astronomical Society of the Pacific, 122, 905.
\bibitem[Kalas, et al.(2004)]{kal04} Kalas, P., Liu, M.~C. \& Matthews, B.~C.\ 2004, Science, 303, 1990.
\bibitem[Kennedy \& Kenyon(2009)]{ken09} Kennedy, G.~M. \& Kenyon, S.~J.\ 2009, \apj, 695, 1210.
\bibitem[Kenworthy \& Mamajek(2015)]{ken15} Kenworthy, M.~A. \& Mamajek, E.~E.\ 2015, \apj, 800, 126.
\bibitem[Kirkpatrick, et al.(2015)]{kir15} Kirkpatrick, A., Pope, A., Sajina, A., et al.\ 2015, \apj, 814, 9.
\bibitem[Klein, et al.(2003)]{kle03} Klein, R., Apai, D., Pascucci, I., et al.\ 2003, \apj, 593, L57.
\bibitem[K{\'o}sp{\'a}l et al.(2013)]{kos13} K{\'o}sp{\'a}l, {\'A}., Mo{\'o}r, A., Juh{\'a}sz, A., et al.\ 2013, \apj, 776, 77.
\bibitem[Kral et al.(2017)]{kra17} Kral, Q., Matr{\`a}, L., Wyatt, M.~C., et al.\ 2017, \mnras, 469, 521.
\bibitem[Laughlin, et al.(2004)]{lau04} Laughlin, G., Bodenheimer, P. \& Adams, F.~C.\ 2004, \apj, 612, L73.
\bibitem[Lestrade, et al.(2006)]{les06} Lestrade, J.-F., Wyatt, M.~C., Bertoldi, F., et al.\ 2006, \aap, 460, 733
\bibitem[Lestrade, et al.(2012)]{les12} Lestrade, J.-F., Matthews, B.~C., Sibthorpe, B., et al.\ 2012, \aap, 548, A86.
\bibitem[Lieman-Sifry, et al.(2016)]{lie16} Lieman-Sifry, J., Hughes, A.~M., Carpenter, J.~M., et al.\ 2016, \apj, 828, 25.
\bibitem[Lindegren et al.(2018)]{lin18} Lindegren, L., Hern{\'a}ndez, J., Bombrun, A., et al.\ 2018, \aap, 616, A2.
\bibitem[Liu, et al.(2004)]{liu04} Liu, M.~C., Matthews, B.~C., Williams, J.~P., et al.\ 2004, \apj, 608, 526.
\bibitem[Lisse et al.(2009)]{lis09} Lisse, C.~M., Chen, C.~H., Wyatt, M.~C., et al.\ 2009, \apj, 701, 2019.
\bibitem[Low, et al.(2005)]{low05} Low, F.~J., Smith, P.~S., Werner, M., et al.\ 2005, \apj, 631, 1170.
\bibitem[Luhman \& Mamajek(2012)]{luh12} Luhman, K.~L. \& Mamajek, E.~E.\ 2012, \apj, 758, 31.
\bibitem[MacGregor, et al.(2013)]{mac13} MacGregor, M.~A., Wilner, D.~J., Rosenfeld, K.~A., et al.\ 2013, \apj, 762, L21.
\bibitem[MacGregor, et al.(2018)]{mac18} MacGregor, M.~A., Weinberger, A.~J., Wilner, D.~J., et al.\ 2018, \apj, 855, L2.
\bibitem[Marino et al.(2016)]{mar16} Marino, S., Matr{\`a}, L., Stark, C., et al.\ 2016, \mnras, 460, 2933.
\bibitem[Matr{\`a}, et al.(2015)]{matra15} Matr{\`a}, L., Pani{\'c}, O., Wyatt, M.~C., et al.\ 2015, \mnras, 447, 3936.
\bibitem[Matr{\`a} et al.(2017)]{mat17} Matr{\`a}, L., MacGregor, M.~A., Kalas, P., et al.\ 2017, \apj, 842, 9.
\bibitem[Matr{\`a}, et al.(2018)]{mat18a} Matr{\`a}, L., Wilner, D.~J., {\"O}berg, K.~I., et al.\ 2018, \apj, 853, 147.
\bibitem[Matr{\`a} et al.(2018)]{mat18} Matr{\`a}, L., Marino, S., Kennedy, G.~M., et al.\ 2018, \apj, 859, 72.
\bibitem[Matthews, et al.(2007)]{mat07} Matthews, B.~C., Kalas, P.~G. \& Wyatt, M.~C.\ 2007, \apj, 663, 1103.
\bibitem[Matthews, et al.(2014)]{mat14} Matthews, B.~C., Krivov, A.~V., Wyatt, M.~C., et al.\ 2014, Protostars and Planets VI, 521.
\bibitem[Matthews, et al.(2015)]{mat15} Matthews, B.~C., Kennedy, G., Sibthorpe, B., et al.\ 2015, \apj, 811, 100.
\bibitem[McMullin, et al.(2007)]{mcm07} McMullin, J.~P., Waters, B., Schiebel, D., et al.\ 2007, Astronomical Data Analysis Software and Systems XVI, 127.
\bibitem[Meng et al.(2012)]{men12} Meng, H.~Y.~A., Rieke, G.~H., Su, K.~Y.~L., et al.\ 2012, \apj, 751, L17.
\bibitem[Meng et al.(2014)]{men14} Meng, H.~Y.~A., Su, K.~Y.~L., Rieke, G.~H., et al.\ 2014, Science, 345, 1032.
\bibitem[Meng, et al.(2015)]{men15} Meng, H.~Y.~A., Su, K.~Y.~L., Rieke, G.~H., et al.\ 2015, \apj, 805, 77.
\bibitem[Mohanty, et al.(2013)]{moh13} Mohanty, S., Greaves, J., Mortlock, D., et al.\ 2013, \apj, 773, 168.
\bibitem[Mo{\'o}r, et al.(2017)]{moo17} Mo{\'o}r, A., Cur{\'e}, M., K{\'o}sp{\'a}l, {\'A}., et al.\ 2017, \apj, 849, 123.
\bibitem[Mulders, et al.(2015)]{mul15} Mulders, G.~D., Pascucci, I. \& Apai, D.\ 2015, \apj, 814, 130.
\bibitem[Murphy, et al.(2018)]{mur18} Murphy, S.~J., Mamajek, E.~E. \& Bell, C.~P.~M.\ 2018, \mnras, 476, 3290.
\bibitem[Olofsson et al.(2018)]{olo18} Olofsson, J., van Holstein, R.~G., Boccaletti, A., et al.\ 2018, \aap, 617, A109.
\bibitem[Payne \& Lodato(2007)]{pay07} Payne, M.~J. \& Lodato, G.\ 2007, \mnras, 381, 1597.
\bibitem[Plavchan, et al.(2005)]{pla05} Plavchan, P., Jura, M. \& Lipscy, S.~J.\ 2005, \apj, 631, 1161.
\bibitem[Plavchan, et al.(2009)]{pla09} Plavchan, P., Werner, M.~W., Chen, C.~H., et al.\ 2009, \apj, 698, 1068.
\bibitem[Reiners(2009)]{rei09} Reiners, A.\ 2009, \apjl, 702, L119 
\bibitem[Ribas, et al.(2015)]{rib15} Ribas, {\'A}., Bouy, H. \& Mer{\'\i}n, B.\ 2015, \aap, 576, A52.
\bibitem[Ricci, et al.(2012)]{ric12} Ricci, L., Testi, L., Natta, A., et al.\ 2012, \apj, 761, L20.
\bibitem[Ricci, et al.(2013)]{ric13} Ricci, L., Isella, A., Carpenter, J.~M., et al.\ 2013, \apj, 764, L27.
\bibitem[Ricci, et al.(2014)]{ric14} Ricci, L., Testi, L., Natta, A., et al.\ 2014, \apj, 791, 20.
\bibitem[Schaefer, et al.(2009)]{sch09} Schaefer, G.~H., Dutrey, A., Guilloteau, S., et al.\ 2009, \apj, 701, 698.
\bibitem[Schneider \& Shkolnik(2018)]{sch18} Schneider, A.~C. \& Shkolnik, E.~L.\ 2018, \aj, 155, 122. 
\bibitem[Sch{\"u}ppler, et al.(2015)]{sch15} Sch{\"u}ppler, C., L{\"o}hne, T., Krivov, A.~V., et al.\ 2015, \aap, 581, A97.
\bibitem[Shangguan, et al.(2018)]{sha18} Shangguan, J., Ho, L.~C. \& Xie, Y.\ 2018, \apj, 854, 158. 
\bibitem[Silverberg et al.(2016)]{sil16} Silverberg, S.~M., Kuchner, M.~J., Wisniewski, J.~P., et al.\ 2016, \apjl, 830, L28
\bibitem[Sissa, et al.(2018)]{sis18} Sissa, E., Olofsson, J., Vigan, A., et al.\ 2018, \aap, 613, L6.
\bibitem[Strubbe \& Chiang(2006)]{str06} Strubbe, L.~E. \& Chiang, E.~I.\ 2006, \apj, 648, 652.
\bibitem[Su, et al.(2009)]{su09} Su, K.~Y.~L., Rieke, G.~H., Stapelfeldt, K.~R., et al.\ 2009, \apj, 705, 314.
\bibitem[Su, et al.(2015)]{su15} Su, K.~Y.~L., Morrison, S., Malhotra, R., et al.\ 2015, \apj, 799, 146.
\bibitem[Su, et al.(2017)]{su17} Su, K.~Y.~L., MacGregor, M.~A., Booth, M., et al.\ 2017, \aj, 154, 225.
\bibitem[Sulentic, et al.(2000)]{sul00} Sulentic, J.~W., Marziani, P. \& Dultzin-Hacyan, D.\ 2000, Annual Review of Astronomy and Astrophysics, 38, 521.
\bibitem[Testi, et al.(2016)]{tes16} Testi, L., Natta, A., Scholz, A., et al.\ 2016, \aap, 593, A111.
\bibitem[van der Plas, et al.(2016)]{van16} van der Plas, G., M{\'e}nard, F., Ward-Duong, K., et al.\ 2016, \apj, 819, 102.
\bibitem[Vidotto \& Bourrier(2017)]{vid17} Vidotto, A.~A. \& Bourrier, V.\ 2017, \mnras, 470, 4026.
\bibitem[Visser, et al.(2009)]{vis09} Visser, R., van Dishoeck, E.~F. \& Black, J.~H.\ 2009, \aap, 503, 323.
\bibitem[White et al.(2016)]{whi16} White, J.~A., Boley, A.~C., Hughes, A.~M., et al.\ 2016, \apj, 829, 6.
\bibitem[Wood(2004)]{woo04} Wood, B.~E.\ 2004, Living Reviews in Solar Physics, 1, 2.
\bibitem[Wyatt, et al.(2015)]{wya15} Wyatt, M.~C., Pani{\'c}, O., Kennedy, G.~M., et al.\ 2015, \apss, 357, 103.

\end{thebibliography}
\end{document}